\documentclass[journal,twocolumn]{IEEEtran}   
\usepackage{graphicx}
\usepackage{epstopdf}
\usepackage{amssymb}
\usepackage{amsfonts}
\usepackage{amsmath}
\usepackage{algorithm}
\usepackage{algorithmic}
\usepackage{subeqnarray}
\usepackage{cases}
\usepackage{bm}
\usepackage{subfigure,amsmath,amssymb,cite}
\usepackage{stfloats}

\usepackage{float}
\usepackage{cuted}
\UseRawInputEncoding

\ifCLASSINFOpdf
\else
\fi
\begin{document}

\title{Beamforming design for Hybrid IRS-aided AF Relay Wireless Networks}

\author{Xuehui~Wang,~Yifan~Zhao,~Feng Shu,~Yan~Wang

\thanks{This work was supported in part by the National Natural Science Foundation of China (Nos.U22A2002, 62071234 and 61972093), the Major Science and Technology plan of Hainan Province under Grant ZDKJ2021022, and the Scientific Research Fund Project of Hainan University under Grant KYQD(ZR)-21008.\emph{(Corresponding authors: Feng Shu)}.}

\thanks{Xuehui Wang, Yifan Zhao and Yan Wang are with the School of Information and Communication Engineering, Hainan University, Haikou, 570228, China.}

\thanks{Feng Shu is with the School of Information and Communication Engineering, Hainan University, Haikou 570228, China, and also with the
School of Electronic and Optical Engineering, Nanjing University of Science and Technology, Nanjing 210094, China (e-mail: shufeng0101@163.com).}

}

\maketitle

\begin{abstract}

In this paper, a hybrid IRS-aided amplify-and-forward (AF) relay wireless network is put forward, where the hybrid IRS is made up of passive and active elements. For maximum signal-to-noise ratio (SNR), a low-complexity method based on successive convex approximation and fractional programming (LC-SCA-FP) is proposed to  jointly optimize the beamforming matrix at AF relay and the reflecting coefficient matrices at IRS. Simulation results verify that the rate achieved by the proposed LC-SCA-FP method surpass those of the benchmark schemes, namely the passive IRS-aided AF relay and only AF relay network.

\end{abstract}


\section{Introduction}
As the Internet-of-Things (IoT) smart devices grow explosively, some existing technologies \cite{2016HRW}, such as massive multiple-input multiple-out (MIMO) and millimeter wave (mmWave) are far from complying with the demands \cite{2020AO}, e.g., massive connectivity, extended coverage, low latency, low cost and low power. Adding relay nodes in the wireless networks can not only save the number of base stations, but also improve the throughput and reliability \cite{2010YE}. However, the relay needs much energy to process signals \cite{2021GQ}. Owing to the benefits of programmability, low circuit cost and low power consumption, intelligent reflecting surface (IRS) is attractive \cite{2020YGH,2020WQQ}. Combining the advantages of IRS and relay is interesting, which can balance cost, power consumption and performance \cite{2022WXH}. In \cite{2020AZ}, compared with only IRS network, it was proved that an IRS-assisted a single-antenna decode-and-forward relay network could harvest the same rate performance with less IRS elements.

However, the received signal via the reflecting channel link experiences large-scale fading twice (i.e., ``double fading'' effect), the received signal is weak in fact \cite{2023ZZJ}. Aiming at eliminating the ``double fading'' effect, the active IRS has emerged as an innovative technology, which needs extra power supply to reflect and amplify the incident signals for obvious performance advancement \cite{2022DLM}. With the same overall power budget, it was proved that the ``double fading'' effect could be broken by active IRS in \cite{2021LRZ}. In addition, when the number of active IRS elements is small-scale or medium-scale, or power budget is sufficient, the active IRS surpassed passive IRS \cite{2022ZKD}.
Given that active IRS can amplify the incident signals and in order to achieve higher rate performance or save more passive IRS elements of the combination network of IRS and relay, we propose a combination network of hybrid IRS and AF relay, which can better balance the circuit cost, power consumption and rate performance. To date, the research work on the combination network is lack. In?the?circumstance,  we derive a low-complexity method to obtain rate improvement or extended coverage range based on maximizing signal-to-noise ratio (SNR). Our contributions are summarized below:

\begin{enumerate}

\item

A combination network composed of hybrid IRS and AF relay is designed, where the hybrid IRS not only includes active IRS elements but also passive IRS elements. Aiming to solve the non-convex problem of maximizing SNR, a low-complexity method using successive convex approximation and fractional programming (LC-SCA-FP) algorithm is presented. Firstly, we approximate the numerator of objective function as a linear function via SCA. Then using Dinkelbach's transformation of FP algorithm and relaxing the non-convex unit-modulus constraint for passive IRS element, the non-convex problem is correspondingly transformed to convex, and can be solved.

\item
The simulation results verify that regardless of the transmit power $P_s$ at source, the transmit power $P_i$ at IRS and the number $K$ of active IRS elements, the proposed LC-SCA-FP method performs much better than the benchmark schemes, such as passive IRS-aided AF relay network, passive IRS-aided AF relay network with random phase and only AF relay network from the rate perspective. Specifically, when $P_s$ $=$ 30dBm or $P_i$ $=$ 40dBm or $K$ $=$ 4, the proposed LC-SCA-FP method can approximately harvest at least 45.0\% rate gain over the benchmark schemes.

\end{enumerate}

Notation: Scalars, vectors and matrices are represented by letters of lower case, bold lower case and bold upper case, respectively. The sign $\textbf{I}_{N}$ is the $N\times N$ identity matrix. Matrix conjugate, transpose and conjugate transpose are respectively denoted as $(\cdot)^*$, $(\cdot)^T$ and $(\cdot)^H$. $| \cdot |$ and $\|\cdot\|$ respectively denote the modulus of a scalar and 2-norm. $\otimes$ and $\odot$ respectively denote Kronecker product and Hadamard product.

\section{System Model}
\begin{figure}[h]
\centering
\includegraphics[width=0.400\textwidth,height=0.200\textheight]{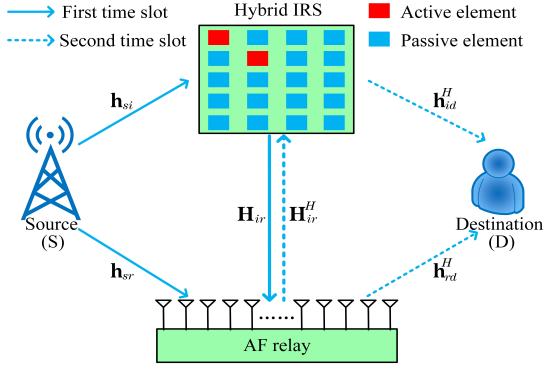}\\
\caption{System model for a hybrid IRS-aided AF relay wireless network.}\label{System_Model.eps}
\end{figure}
A hybrid IRS-aided AF relay network is sketched in Fig. 1, where the network is operated in a time division half-duplex scenario. Source (S) and destination (D) are with a single antenna, respectively. A AF relay is with $M$ antennas, and a hybrid IRS includes $N$ elements consisting of $K$ active elements and $L$ passive elements, i.e., $N=K+L$. Let us define $\cal E_N$, $\cal E_K$ and $\cal E_L$ as the sets of $N$ elements, $K$ active elements and $L$ passive elements, respectively. Furthermore, $\cal E_N=\cal E_K\cup\cal E_L$ and $\cal E_K\cap\cal E_L=\varnothing$. The reflecting coefficient matrices of $\cal E_N$, $\cal E_K$ and $\cal E_L$ are respectively denoted by $\boldsymbol\Theta$, $\boldsymbol\Phi$ and $\boldsymbol\Psi$, where $\boldsymbol\Theta = \text{diag}( \alpha_1, \cdots,\alpha_N )$, and the reflecting coefficients of $i$th element in $\boldsymbol\Theta$ is expressed by
\begin{subnumcases}{\alpha_i=}{}
|\beta_i| e^{j\theta _{i}}~~~~~~~~~i\in \cal E_K, \\
e^{j\theta _{i}}~~~~~~~~~~~~~i\in \cal E_L,
\end{subnumcases}
where $|\beta_i|$ and $\theta _{i}\in ( 0 ,{2\pi } ]$ are amplifying coefficient and phase shift of the $i$th element.
For convenience of derivation below, we have
$\boldsymbol\Theta=\boldsymbol\Phi+\boldsymbol\Psi$, $\boldsymbol\Phi=\textbf{E}_K\boldsymbol\Theta$, $\boldsymbol\Psi=\overline{\textbf{E}}_K\boldsymbol\Theta$,
where $\textbf{E}_K+\overline{\textbf{E}}_K=\textbf{I}_N$ and $\textbf{E}_K\overline{\textbf{E}}_K=\textbf{0}_N$.
$\boldsymbol\Phi$, $\boldsymbol\Psi$, $\textbf{E}_K$ and $\overline{\textbf{E}}_K$ are sparse diagonal matrices. The $k$th non-zero value of the diagonal corresponding to the $k$th active element is 1, thus there are $K$ values being 1 and the rest $L$ values being 0 on the diagonal of $\textbf{E}_K$. Additionally, $\overline{\textbf{E}}_K$ is similar to $\textbf{E}_K$.
It is assumed that there are obstacles in the direct link between S and D, which results the direct signal can be ignored. While the power of signals reflected by the IRS twice or more are so weak that they can also be ignored. In the first time slot, the received signal at IRS can be denoted as
\begin{equation}\label{y_1i_r}
\textbf{y}_{1i}^r=\sqrt {{P_s}}\textbf{h}_{si}x+\textbf{n}_{1i},
\end{equation}
where $x$ with $\mathbb{E}\{x^H{x}\}=1$ is the signal transmitted from S, and $P_s$ is the corresponding transmit power.
We assume all channels follow Rayleigh fading, $\textbf{h}_{si}\in \mathbb C^{N \times 1}$ is the channel from S to IRS. $\textbf{n}_{1i}$ represents the additive white Gaussian noise (AWGN) at IRS with distribution $\textbf{n}_{1i}\sim \mathcal{CN}( \textbf{0},\sigma_{1i}^2\textbf{E}_K{\bf I}_{N} )$, which is caused by $K$ active elements. The received signal at AF relay is given by
\begin{equation}\label{y_r}
\textbf{y}_r=\sqrt {P_s}(\textbf{h}_{sr} + \textbf{H}_{ir}\boldsymbol\Theta_1\textbf{h}_{si})x + \textbf{H}_{ir}\textbf{E}_K\boldsymbol\Theta_1\textbf{n}_{1i} + \textbf{n}_r,
\end{equation}
where $\textbf{h}_{sr}\in \mathbb C^{M \times 1}$ and $\textbf{H}_{ir}\in \mathbb C^{M \times N}$ are the channels from S to AF relay and IRS to AF relay. $\boldsymbol\Theta_1 = \text{diag}( \alpha_{11}, \cdots,\alpha_{1N} )$ is the reflecting coefficient matrix of $\cal E_N$ in the first time slot. $\textbf{n}_r\sim \mathcal{CN}( \textbf{0},\sigma_r^2{\bf I}_{M} )$ is the AWGN at AF relay.
After performing receive and transmit beamforming, the transmit signal at AF relay can be expressed as
\begin{equation}\label{y_t}
\textbf{y}_t=\textbf{A}\textbf{y}_r,
\end{equation}
where $ \textbf{A}\in \mathbb C^{M \times M} $ is the beamforming matrix. In the second time slot, the received signal at IRS is written by
\begin{equation}\label{y_2i}
\textbf{y}_{2i}^r=\textbf{H}_{ir}^H\textbf{y}_t+\textbf{n}_{2i},
\end{equation}
where $ \textbf{H}_{ir}^H\in \mathbb C^{N \times M} $ is the channel from AF relay to IRS. $\textbf{n}_{2i}\sim \mathcal{CN}( \textbf{0},\sigma_{2i}^2\textbf{E}_K{\bf I}_{N} )$ is the noise. The received signal at D is as follows
\begin{equation}\label{y_d}
y_d=(\textbf{h}_{rd}^H + \textbf{h}_{id}^H\boldsymbol\Theta_2\textbf{H}_{ir}^H)\textbf{y}_t + \textbf{h}_{id}^H\textbf{E}_K\boldsymbol\Theta_2\textbf{n}_{2i} + \text{n}_d,
\end{equation}
where $\textbf{h}_{rd}^H\in \mathbb C^{1 \times M}$ and $\textbf{h}_{id}^H\in \mathbb C^{1 \times N}$ are the channels from AF relay to D and IRS to D. $\boldsymbol\Theta_2 = \text{diag}( \alpha_{21}, \cdots,\alpha_{2N} )$ is the reflecting coefficient matrix of $\cal E_N$ in the second time slot. $\text{n}_d\sim \mathcal{CN}( 0,\sigma_d^2 )$ is the AWGN at D. Substituting (\ref{y_r}) and (\ref{y_t}) into (\ref{y_d}) yields
\begin{align}\label{y_d1}
&y_d=\sqrt {P_s}(\textbf{h}_{rd}^H + \textbf{h}_{id}^H\boldsymbol\Theta_2\textbf{H}_{ir}^H)\textbf{A}(\textbf{h}_{sr} +
\textbf{H}_{ir}\boldsymbol\Theta_1\textbf{h}_{si})x \nonumber\\
&~~~~~~~+ (\textbf{h}_{rd}^H + \textbf{h}_{id}^H\boldsymbol\Theta_2\textbf{H}_{ir}^H)\textbf{A}( \textbf{H}_{ir}\textbf{E}_K\boldsymbol\Theta_1\textbf{n}_{1i} + \textbf{n}_r ) \nonumber\\
&~~~~~~~+ \textbf{h}_{id}^H\textbf{E}_K\boldsymbol\Theta_2\textbf{n}_{2i} + \text{n}_d.
\end{align}
It is assumed that $ \sigma_{1i}^2=\sigma_{2i}^2=\sigma_{r}^2=\sigma_d^2=\sigma^2 $ and $\gamma_s=\frac{P_s}{\sigma^2} $, the achievable system rate can be defined as
\begin{equation}\label{R}
R=\frac{1}{2}\log_2 (1 + \text{SNR}),
\end{equation}
where SNR can be formulated as (\ref{SNR_1}), as shown at the top of next page.
 \begin{figure*}[ht] 
 	\centering
      \begin{equation}\label{SNR_1}
        \text{SNR}=\frac{\gamma_s|(\textbf{h}_{rd}^H + \textbf{h}_{id}^H\boldsymbol\Theta_2\textbf{H}_{ir}^H)\textbf{A}(\textbf{h}_{sr} +
          \textbf{H}_{ir}\boldsymbol\Theta_1\textbf{h}_{si})|^2}  {\|(\textbf{h}_{rd}^H + \textbf{h}_{id}^H\boldsymbol\Theta_2\textbf{H}_{ir}^H)\textbf{A}\textbf{H}_{ir}\textbf{E}_K\boldsymbol\Theta_1\|^2+ \|(\textbf{h}_{rd}^H + \textbf{h}_{id}^H\boldsymbol\Theta_2\textbf{H}_{ir}^H)\textbf{A}\|^2+ \|\textbf{h}_{id}^H\textbf{E}_K\boldsymbol\Theta_2\|^2 + 1}. \nonumber
      \end{equation}
  \hrulefill
 \end{figure*}

To enhance the system rate performance, it is necessary to maximize SNR due to the log function is a monotone increasing function of SNR. The optimization problem can be cast as
\begin{subequations}\label{OP}
\begin{align}
&\max \limits_{\boldsymbol\Theta_1, \boldsymbol\Theta_2, \textbf{A} } \text{SNR} \\
&~~~\text{s.t.}~~   |\boldsymbol\Theta_1(i,i)|=1,~|\boldsymbol\Theta_2(i,i)|=1,~\text{for}~i\in \cal E_L,  \label{OP_4}\\
&~~~~~~~~  \gamma_s\|\textbf{E}_K\boldsymbol\Theta_1\textbf{h}_{si}\|^2+\|\textbf{E}_K\boldsymbol\Theta_1\|_F^2\leq \gamma_i,\label{OP_1}\\
&~~~~~~~~  \gamma_s\|\textbf{A}(\textbf{h}_{sr} + \textbf{H}_{ir}\boldsymbol\Theta_1\textbf{h}_{si})\|^2 \nonumber\\
&~~~~~~~~~~~~~~~~~  + \|\textbf{A}\textbf{H}_{ir}\textbf{E}_K\boldsymbol\Theta_1\|_F^2+ \|\textbf{A}\|_F^2\leq \gamma_r,\label{OP_2}\\
&~~~~~~~~  \gamma_s\|\textbf{E}_K\boldsymbol\Theta_2\textbf{H}_{ir}^H\textbf{A}(\textbf{h}_{sr} +\textbf{H}_{ir}\boldsymbol\Theta_1\textbf{h}_{si})\|^2 \nonumber\\
&~~~~~~~~~~~~~~~~~  + \|\textbf{E}_K\boldsymbol\Theta_2\textbf{H}_{ir}^H\textbf{A}\textbf{H}_{ir}\textbf{E}_K\boldsymbol\Theta_1\|_F^2 \nonumber\\
&~~~~~~~~~~~~~~~~~  + \|\textbf{E}_K\boldsymbol\Theta_2\textbf{H}_{ir}^H\textbf{A}\|_F^2+\|\textbf{E}_K\boldsymbol\Theta_2\|_F^2\leq \gamma_i,  \label{OP_3}
\end{align}
\end{subequations}
where $\gamma_i=\frac{P_i}{\sigma^2} $ and $\gamma_r=\frac{P_r}{\sigma^2} $, $P_i$ and $P_r$ respectively denote the transmit power budgets at IRS and AF relay.

\section{Proposed a Low-complexity SCA-FP-based Max-SNR Method}
\subsection{Optimize $\textbf{A}$ With Fixed $\boldsymbol\Theta_1$ and $\boldsymbol\Theta_2$}
Given $\boldsymbol\Theta_1$ and $\boldsymbol\Theta_2$, the optimization problem is reduced to
\begin{subequations}\label{OP_1}
\begin{align}
&\max \limits_{ \textbf{A} } ~~~~\text{SNR}~~~~~~~~~~~   \\ 
&~\text{s.t.}~~~  \gamma_s\|\textbf{A}(\textbf{h}_{sr} + \textbf{H}_{ir}\boldsymbol\Theta_1\textbf{h}_{si})\|^2 \nonumber\\
&~~~~~~~~~~~~~~~~~  + \|\textbf{A}\textbf{H}_{ir}\textbf{E}_K\boldsymbol\Theta_1\|_F^2+ \|\textbf{A}\|_F^2\leq \gamma_r, \\
&~~~~~~~~  \gamma_s\|\textbf{E}_K\boldsymbol\Theta_2\textbf{H}_{ir}^H\textbf{A}(\textbf{h}_{sr} +\textbf{H}_{ir}\boldsymbol\Theta_1\textbf{h}_{si})\|^2 \nonumber\\
&~~~~~~~~~~~~~~~~~  + \|\textbf{E}_K\boldsymbol\Theta_2\textbf{H}_{ir}^H\textbf{A}\textbf{H}_{ir}\textbf{E}_K\boldsymbol\Theta_1\|_F^2 \nonumber\\
&~~~~~~~~~~~~~~~~~  + \|\textbf{E}_K\boldsymbol\Theta_2\textbf{H}_{ir}^H\textbf{A}\|_F^2+\|\textbf{E}_K\boldsymbol\Theta_2\|_F^2\leq \gamma_i.
\end{align}
\end{subequations}
Defining $\textbf{a}=\text{vec}(\textbf{A})\in \mathbb C^{M^2 \times 1}$,  SNR can be translated to
\begin{equation}\label{SNR_2}
\text{SNR}=\frac{ \gamma_s\textbf{a}^H\textbf{B}_1\textbf{a} } {\textbf{a}^H(\textbf{B}_2+\textbf{B}_3)\textbf{a}+  \|\textbf{h}_{id}^H\textbf{E}_K\boldsymbol\Theta_2\|^2 + 1 },
\end{equation}
where
$\textbf{B}_1=[(\textbf{h}_{sr} + \textbf{H}_{ir}\boldsymbol\Theta_1\textbf{h}_{si})^*(\textbf{h}_{sr} + \textbf{H}_{ir}\boldsymbol\Theta_1\textbf{h}_{si})^T]\otimes[(\textbf{h}_{rd}^H + \textbf{h}_{id}^H\boldsymbol\Theta_2\textbf{H}_{ir}^H)^H(\textbf{h}_{rd}^H + \textbf{h}_{id}^H\boldsymbol\Theta_2\textbf{H}_{ir}^H)]$,
$\textbf{B}_2=[(\textbf{H}_{ir}\textbf{E}_K\boldsymbol\Theta_1)^*(\textbf{H}_{ir}\textbf{E}_K\boldsymbol\Theta_1)^T]\otimes[(\textbf{h}_{rd}^H + \textbf{h}_{id}^H\boldsymbol\Theta_2\textbf{H}_{ir}^H)^H(\textbf{h}_{rd}^H + \textbf{h}_{id}^H\boldsymbol\Theta_2\textbf{H}_{ir}^H)]$,
$\textbf{B}_3=\textbf{I}_{M}\otimes[(\textbf{h}_{rd}^H + \textbf{h}_{id}^H\boldsymbol\Theta_2\textbf{H}_{ir}^H)^H(\textbf{h}_{rd}^H + \textbf{h}_{id}^H\boldsymbol\Theta_2\textbf{H}_{ir}^H)]$.
In the same manner, the constraints of problem (\ref{OP_1}) can be respectively converted to
\begin{subequations}
\begin{align}
&\textbf{a}^H(\gamma_s\textbf{C}_1+\textbf{C}_2+\textbf{I}_{M^2})\textbf{a}\leq\gamma_r,\label{OP_2_1} \\
&\textbf{a}^H(\gamma_s\textbf{D}_1+\textbf{D}_2+\textbf{D}_3)\textbf{a}+\|\textbf{E}_K\boldsymbol\Theta_2\|_F^2\leq \gamma_i,\label{OP_3_1}
\end{align}
\end{subequations}
where
$\textbf{C}_1=[(\textbf{h}_{sr} + \textbf{H}_{ir}\boldsymbol\Theta_1\textbf{h}_{si})^*(\textbf{h}_{sr} + \textbf{H}_{ir}\boldsymbol\Theta_1\textbf{h}_{si})^T]\otimes\textbf{I}_M$,
$\textbf{C}_2=[(\textbf{H}_{ir}\textbf{E}_K\boldsymbol\Theta_1)^*(\textbf{H}_{ir}\textbf{E}_K\boldsymbol\Theta_1)^T]\otimes\textbf{I}_M$,
$\textbf{D}_1=[(\textbf{h}_{sr} + \textbf{H}_{ir}\boldsymbol\Theta_1\textbf{h}_{si})^*(\textbf{h}_{sr} + \textbf{H}_{ir}\boldsymbol\Theta_1\textbf{h}_{si})^T]\otimes[(\textbf{E}_K\boldsymbol\Theta_2\textbf{H}_{ir}^H)^H(\textbf{E}_K
\boldsymbol\Theta_2\textbf{H}_{ir}^H)]$,
$\textbf{D}_2=[(\textbf{H}_{ir}\textbf{E}_K\boldsymbol\Theta_1)^*(\textbf{H}_{ir}\textbf{E}_K\boldsymbol\Theta_1)^T]\otimes
[(\textbf{E}_K\boldsymbol\Theta_2\textbf{H}_{ir}^H)^H(\textbf{E}_K\boldsymbol\Theta_2\textbf{H}_{ir}^H)]$,
$\textbf{D}_3=\textbf{I}_M\otimes[(\textbf{E}_K\boldsymbol\Theta_2\textbf{H}_{ir}^H)^H(\textbf{E}_K\boldsymbol\Theta_2\textbf{H}_{ir}^H)]$.
The problem (\ref{OP_1}) can be rewritten as
\begin{subequations}\label{a}
\begin{align}
&\max \limits_{ \textbf{a}} ~~~\text{(\ref{SNR_2})} \label{a_1}\\
&~\text{s.t.}~~~~~  \text{(\ref{OP_2_1})},~\text{(\ref{OP_3_1})}.
\end{align}
\end{subequations}
Because of the convex numerator of the objective function, (\ref{a}) is a convex-convex FP problem. It is necessary to convert the convex numerator into a concave function by using SCA method. We approximate the numerator by using a linear function, i.e., its first-order Taylor expansion at feasible vector $\widetilde{\textbf{a}}$ is $\gamma_s\textbf{a}^H\textbf{B}_1\textbf{a} \geq 2\gamma_s\Re\{\textbf{a}^H\textbf{B}_1\widetilde{\textbf{a}}\}-\gamma_s\widetilde{\textbf{a}}^H\textbf{B}_1\widetilde{\textbf{a}}$, where  $\widetilde{\textbf{a}}$ is the solution of previous iteration. Inserting the low bound of $\gamma_s\textbf{a}^H\textbf{B}_1\textbf{a}$ back into problem (\ref{a}) yields
\begin{subequations}
\begin{align}
&\max \limits_{ \textbf{a}} ~~~\frac{ 2\gamma_s\Re\{\textbf{a}^H\textbf{B}_1\widetilde{\textbf{a}}\}-\gamma_s\widetilde{\textbf{a}}^H\textbf{B}_1\widetilde{\textbf{a}} } {\textbf{a}^H(\textbf{B}_2+\textbf{B}_3)\textbf{a}+  \|\textbf{h}_{id}^H\textbf{E}_K\boldsymbol\Theta_2\|^2 + 1 } \\
&~\text{s.t.}~~~~~  \text{(\ref{OP_2_1})},~\text{(\ref{OP_3_1})},
\end{align}
\end{subequations}
which is a concave-convex FP problem. Here, introducing Dinkelbachs transformation to solve the above problem as follows
\begin{subequations}\label{a2}
\begin{align}
&\max \limits_{ \textbf{a}} ~~~2\gamma_s\Re\{\textbf{a}^H\textbf{B}_1\widetilde{\textbf{a}}\}-\gamma_s\widetilde{\textbf{a}}^H\textbf{B}_1\widetilde{\textbf{a}} \nonumber\\
&~~~~~~~~~-\mu[\textbf{a}^H(\textbf{B}_2+\textbf{B}_3)\textbf{a}+  \|\textbf{h}_{id}^H\textbf{E}_K\boldsymbol\Theta_2\|^2 + 1] \label{a2_1}\\
&~\text{s.t.}~~~~~  \text{(\ref{OP_2_1})},~\text{(\ref{OP_3_1})},
\end{align}
\end{subequations}
where $\mu$ is a slack variable and is iteratively updated by
$\mu(t+1)=\frac{ 2\gamma_s\Re\{\textbf{a}^H(t)\textbf{B}_1\widetilde{\textbf{a}}\}-\gamma_s\widetilde{\textbf{a}}^H\textbf{B}_1\widetilde{\textbf{a}} } {\textbf{a}^H(t)(\textbf{B}_2+\textbf{B}_3)\textbf{a}(t)+  \|\textbf{h}_{id}^H\textbf{E}_K\boldsymbol\Theta_2\|^2 + 1}$,
where $t$ is the iteration number. $\mu$ is nondecreasing after each iteration, which guarantees the convergence of the objective function.
Here, problem (\ref{a2}) is a convex optimization problem. When $\widetilde{\textbf{a}}$ and $\mu$ are fixed, $\textbf{a}$ can be directly achieved by CVX. Correspondingly, $\textbf{A}$ can be obtained.

\subsection{Optimize $\boldsymbol\Theta_1$ With Fixed $\textbf{A}$ and $\boldsymbol\Theta_2$}
It is assumed that AF relay beamforming matrix $\textbf{A}$ and $\boldsymbol\Theta_2$ are given. Let us define $\textbf{u}_1=[ \alpha_{11}, \cdots, \alpha_{1N}]^T$, the constraint (\ref{OP_4}) can be rewritten as
\begin{equation}
\left|\textbf{u}_1(i)\right|^2=1,~~\text{for}~i\in \cal E_L, \label{OPu11_1}
\end{equation}
which can be relaxed as
\begin{equation}
\textbf{u}_1^H(i)\textbf{u}_1(i)\leq1,~~\text{for}~i\in \cal E_L. \label{OPu11_1_1}
\end{equation}
Problem (\ref{OP}) with respect to $\textbf{u}_1$ can be rearranged as
\begin{subequations}\label{OPu1_1}
\begin{align}
&\max \limits_{\textbf{u}_1 } ~~\frac{\gamma_s|\textbf{h}_1^H\textbf{u}_1+a |^2} {\|\text{diag}\{\textbf{h}_{rid}^H\textbf{H}_{ir}\textbf{E}_K\}\textbf{u}_1\|^2+ b} \label{OPu1_1_1}\\
&~\text{s.t.}~~~~  \text(\ref{OPu11_1_1}),~ \gamma_s\|\textbf{E}_K\text{diag}\{\textbf{h}_{si}\}\textbf{u}_{1}\|^2  + \|\textbf{E}_K\textbf{u}_{1}\|^2\leq \gamma_i, \label{OPu1_1_3}\\
&~~~~~~~~ \gamma_s\|\textbf{A}\textbf{h}_{sr} + \textbf{P}_1\textbf{u}_{1}\|^2+ \|\textbf{A}\textbf{H}_{ir}\textbf{E}_K\text{diag}\{\textbf{u}_1\}\|_F^2\nonumber\\
&~~~~~~~~ +\|\textbf{A}\|_F^2\leq \gamma_r, \label{OPu1_1_4}\\
&~~~~~~~~ \gamma_s\|\textbf{h}_2+\textbf{P}_2\textbf{u}_{1}\|^2 +\|\textbf{P}_3\textbf{E}_K\text{diag}\{\textbf{u}_1\}\|_F^2\leq\widetilde{\gamma}_i, \label{OPu1_1_5}
\end{align}
\end{subequations}
where
$\textbf{h}_{1}=[(\textbf{h}_{rd}^H + \textbf{h}_{id}^H\boldsymbol\Theta_2\textbf{H}_{ir}^H)\textbf{A}\textbf{H}_{ir}\text{diag}\{\textbf{h}_{si}\}]^H$,
$\textbf{h}_2=\textbf{E}_K\boldsymbol\Theta_2\textbf{H}_{ir}^H\textbf{A}\textbf{h}_{sr},~\textbf{P}_1=\textbf{A}\textbf{H}_{ir}\text{diag}\{\textbf{h}_{si}\}$,
$\textbf{P}_2=\textbf{E}_K\boldsymbol\Theta_2\textbf{H}_{ir}^H\textbf{A}\textbf{H}_{ir}\text{diag}\{\textbf{h}_{si}\}$,
$\textbf{P}_3=\textbf{E}_K\boldsymbol\Theta_2\textbf{H}_{ir}^H\textbf{A}\textbf{H}_{ir},~a=(\textbf{h}_{rd}^H + \textbf{h}_{id}^H\boldsymbol\Theta_2\textbf{H}_{ir}^H)\textbf{A}\textbf{h}_{sr}$,
$b=\|(\textbf{h}_{rd}^H + \textbf{h}_{id}^H\boldsymbol\Theta_2\textbf{H}_{ir}^H)\textbf{A}\|^2+\|\textbf{h}_{id}^H\textbf{E}_K\boldsymbol\Theta_2\|^2 + 1$,
$\widetilde{\gamma}_i=\gamma_i-\|\textbf{E}_K\boldsymbol\Theta_2\textbf{H}_{ir}^H\textbf{A}\|_F^2-\|\textbf{E}_K\boldsymbol\Theta_2\|_F^2$.
Similarly, aiming at converting the numerator of objective function to concave, the first-order Taylor expansion at the point $\widetilde{\textbf{u}}_1$ is employed to $\gamma_s|\textbf{h}_1^H\textbf{u}_1+a |^2$ and transform it into the linear function, i.e.
$|\textbf{h}_1^H\textbf{u}_1+a |^2 \geq 2\Re\{\textbf{u}_1^H\textbf{h}_1 (\textbf{h}_1^H\widetilde{\textbf{u}}_1+a )\}
+a^*a-\widetilde{\textbf{u}}_1^H\textbf{h}_1\textbf{h}_1^H\widetilde{\textbf{u}}_1$.
In the same manner, problem (\ref{OPu1_1}) can be written as
\begin{subequations}\label{OPu1_3}
\begin{align}
&\max \limits_{\textbf{u}_1}~ 2\gamma_s\Re\{\textbf{u}_1^H\textbf{h}_1 (\textbf{h}_1^H\widetilde{\textbf{u}}_1+a )\}
+\gamma_sa^*a-\gamma_s\widetilde{\textbf{u}}_1^H\textbf{h}_1\textbf{h}_1^H\widetilde{\textbf{u}}_1- \nonumber\\
&~~~~~ \omega(b+\textbf{u}_1^H\text{diag}\{\textbf{E}_K\textbf{H}_{ir}^H\textbf{h}_{rid}\}\text{diag}\{\textbf{h}_{rid}^H\textbf{H}_{ir}\textbf{E}_K\}\textbf{u}_1 ) \label{OPu1_3_1}\\
&~\text{s.t.}~~~~  \text(\ref{OPu11_1_1}),~ \gamma_s\|\textbf{E}_K\text{diag}\{\textbf{h}_{si}\}\textbf{u}_{1}\|^2  + \|\textbf{E}_K\textbf{u}_{1}\|^2\leq \gamma_i, \\
&~~~~~~~~ \gamma_s\|\textbf{A}\textbf{h}_{sr} + \textbf{P}_1\textbf{u}_{1}\|^2+ \|\textbf{A}\textbf{H}_{ir}\textbf{E}_K\text{diag}\{\textbf{u}_1\}\|_F^2  \nonumber\\
&~~~~~~~~ +\|\textbf{A}\|_F^2\leq \gamma_r, \\
&~~~~~~~~ \gamma_s\|\textbf{h}_2+\textbf{P}_2\textbf{u}_{1}\|^2 +\|\textbf{P}_3\textbf{E}_K\text{diag}\{\textbf{u}_1\}\|_F^2\leq\widetilde{\gamma}_i, 
\end{align}
\end{subequations}
where $\omega$ is a variable scalar,
\begin{align}\label{w}
&\omega(t+1)=  \nonumber\\
&\frac{2\gamma_s\Re\{\textbf{u}_1^H(t)\textbf{h}_1 (\textbf{h}_1^H\widetilde{\textbf{u}}_1+a )\}
+\gamma_sa^*a-\gamma_s\widetilde{\textbf{u}}_1^H\textbf{h}_1\textbf{h}_1^H\widetilde{\textbf{u}}_1}
{\|\text{diag}\{\textbf{h}_{rid}^H\textbf{H}_{ir}\textbf{E}_K\}\textbf{u}_1(t)\|^2+ b}.
\end{align}
Because of the concave object function and the convex constraints, problem (\ref{OPu1_3}) is convex. For a given feasible vector $\widetilde{\textbf{u}}_1$ and $\omega$,  problem (\ref{OPu1_3}) can be solved by CVX directly, thereby $\textbf{u}_1$ is achieved.


\subsection{Optimize $\boldsymbol\Theta_2$ With Fixed $\textbf{A}$ and $\boldsymbol\Theta_1$}
Similarly, given AF relay beamforming matrix $\textbf{A}$ and $\boldsymbol\Theta_1$, defining
$\textbf{u}_2=[ \alpha_{21}, \cdots, \alpha_{2N}]^H$, the problem related to $\textbf{u}_2$ can be formulated as 
\begin{subequations}\label{SPv2}
\begin{align}
&\max \limits_{\textbf{u}_2}~~ 2\gamma_s\Re\{\textbf{u}_2^H\textbf{h}_3 (\textbf{h}_3^H\widetilde{\textbf{u}}_2+c^* )\}
+\gamma_scc^*-\gamma_s\widetilde{\textbf{u}}_2^H\textbf{h}_3\textbf{h}_3^H\widetilde{\textbf{u}}_2 \nonumber\\
&~~~~~~~~
-\lambda( \textbf{u}_2^H\textbf{Q}_1\textbf{Q}_1^H\textbf{u}_2+2\Re\{\textbf{u}_2^H\textbf{Q}_1\textbf{h}_4\}+\textbf{h}_4^H\textbf{h}_4 ) \nonumber\\
&~~~~~~~~ -\lambda( \textbf{u}_2^H\textbf{Q}_2\textbf{Q}_2^H\textbf{u}_2+2\Re\{\textbf{u}_2^H\textbf{Q}_2\textbf{A}^H\textbf{h}_{rd}\})
 \nonumber\\
&~~~~~~~~ -\lambda\textbf{h}_{rd}^H\textbf{A}\textbf{A}^H\textbf{h}_{rd}-\lambda\textbf{u}_2^H\textbf{Q}_3\textbf{Q}_3^H\textbf{u}_2  -\lambda \label{SPv2_1}\\
&~\text{s.t.}~~~ \textbf{u}_2^H(i)\textbf{u}_2(i)\leq1,~\text{for}~i\in \cal E_L, \\
&~~~~~~~ \textbf{u}_2^H[\gamma_s\textbf{H}_3^H\textbf{H}_3+( \textbf{H}_4\textbf{H}_4^H+\textbf{H}_{ir}^H\textbf{A}\textbf{A}^H\textbf{H}_{ir}+\textbf{I}_N )\odot\textbf{E}_K]\textbf{u}_2 \nonumber\\
&~~~~~~~ \leq\gamma_i,
\end{align}
\end{subequations}
where $\widetilde{\textbf{u}}_2$ is a feasible vector, and
$\textbf{h}_3=\text{diag}\{\textbf{h}_{id}^H\}\textbf{H}_{ir}^H\textbf{A}(\textbf{h}_{sr}+\textbf{H}_{ir}\boldsymbol\Theta_1\textbf{h}_{si})$,
$\textbf{h}_4=(\textbf{h}_{rd}^H\textbf{A}\textbf{H}_{ir}\textbf{E}_K\boldsymbol\Theta_1)^H$,
$\textbf{Q}_1=\text{diag}\{\textbf{h}_{id}^H\}\textbf{H}_{ir}^H\textbf{A}\textbf{H}_{ir}\textbf{E}_K\boldsymbol\Theta_1$,
$\textbf{Q}_2=\text{diag}\{\textbf{h}_{id}^H\}\textbf{H}_{ir}^H\textbf{A},~\textbf{Q}_3=\text{diag}\{\textbf{h}_{id}^H\textbf{E}_K\}$,
$c=\textbf{h}_{rd}^H\textbf{A}(\textbf{h}_{sr}+\textbf{H}_{ir}\boldsymbol\Theta_1\textbf{h}_{si})$,
$\lambda$ is a variable,
$\lambda(t+1)=
\frac{2\gamma_s\Re\{\textbf{u}_2^H(t)\textbf{h}_3 (\textbf{h}_3^H\widetilde{\textbf{u}}_2+c^* )\}
+\gamma_scc^*-\gamma_s\widetilde{\textbf{u}}_2^H\textbf{h}_3\textbf{h}_3^H\widetilde{\textbf{u}}_2}  {\|\textbf{u}_2 ^H(t)\textbf{Q}_1+\textbf{h}_4^H\|^2+\|\textbf{u}_2 ^H(t)\textbf{Q}_2+\textbf{h}_{rd}^H\textbf{A}\|^2+\|\textbf{u}_2 ^H(t)\textbf{Q}_3 \|^2+1}$. 
Given $\widetilde{\textbf{u}}_2$ and $\lambda$, problem (\ref{SPv2}) can be effectively solved by CVX, while $\textbf{u}_2$ can be obtained.

\section{Simulation and Numerical Results}
In this section, we assume S, D, hybrid IRS and AF relay are located in three-dimensional (3D) space, where S, D, hybrid IRS and AF relay are located at (0, 0, 0), (0, 100, 0), ($-$10, 50, 20) and (10, 50, 10) in meter (m), respectively. The path loss is calculated in line with $PL(d)=PL_0-10{\alpha}\text{log}_{10}(\frac{d}{d_0})$, where $PL_0= -$30dB is the path loss at the reference distance $d_0=1$m, $d$ is the distance between transmitter and receiver, and $\alpha$ is the path loss exponent. Here, the path loss exponents of each channel link S-IRS, IRS-AF relay, IRS-D, S-AF relay and AF relay-D are set as 2.0, 2.0, 2.0, 3.0 and 3.0, respectively. Additionally, we set $\sigma^2= -$80dBm and $\textbf{E}_K$ is randomly generated.

Here, passive IRS-aided AF relay network, passive IRS-aided AF relay network with random phase and only AF relay network are regarded as the benchmark schemes. To fairly compare the benchmark schemes with our proposed network, we set the power budgets of AF relay of the benchmark schemes is equal to that of the  proposed  hybrid IRS and AF relay network, i.e., $P_R=P_i+P_r$.

\begin{figure}[h]
\centering
\includegraphics[width=0.350\textwidth,height=0.200\textheight]{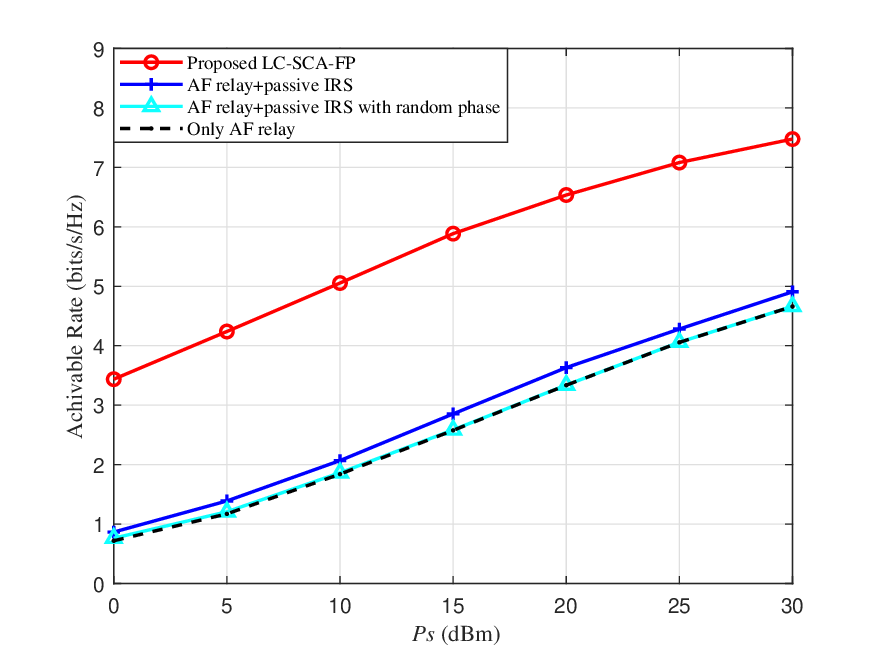}\\
\caption{  Achievable rate versus $P_s$ with $(M, N, K)$ $= (2, 32, 4)$.}\label{figure5_Rate_Vs_Ps}
\end{figure}
\begin{figure}[h]
\centering
\includegraphics[width=0.350\textwidth,height=0.200\textheight]{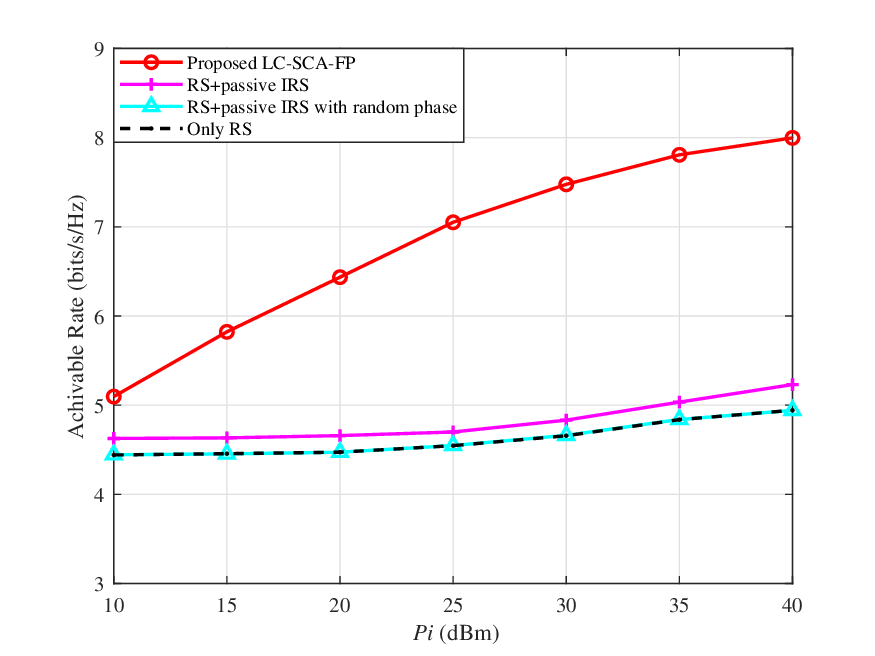}\\
\caption{  Achievable rate versus $P_i$ with $(M, N, K, P_s)$ $=$ (2, 32, 4, 30dBm).}\label{figure6_Rate_Vs_Pi}
\end{figure}
\begin{figure}[h]
\centering
\includegraphics[width=0.350\textwidth,height=0.200\textheight]{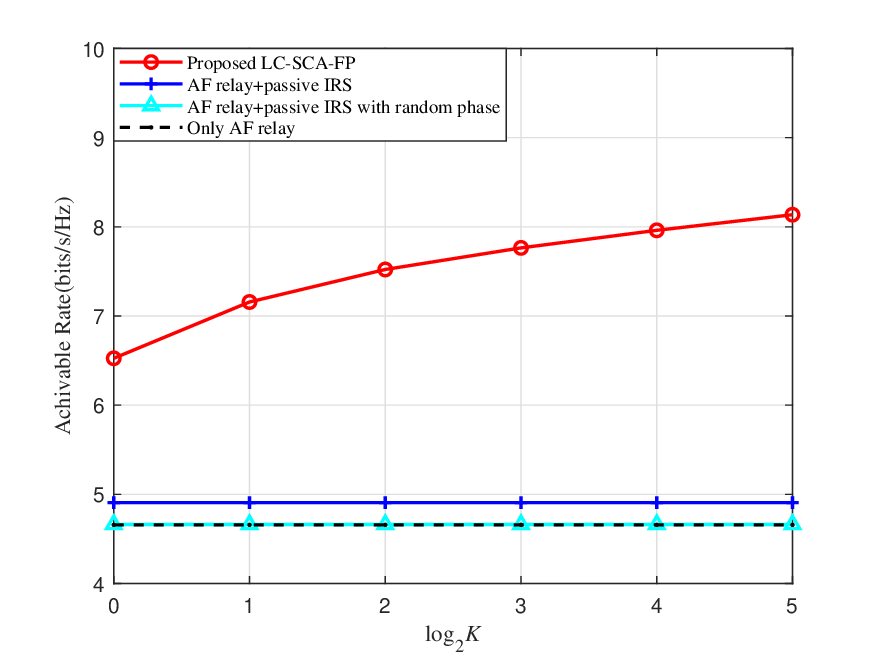}\\
\caption{  Achievable rate versus $K$ with $(M, N, P_s)$ $=$ (2, 32, 30dBm).}\label{figure7_Rate_Vs_K}
\end{figure}

Fig. 2 shows the achievable rate versus $P_s$ with $(M, N, K)$ $=$ (2, 32, 4). It can be seen that the proposed LC-SCA-FP method with $(P_i, P_r)$ $=$ (30dBm, 30dBm) perform better than AF relay+passive IRS, AF relay+passive IRS with random phase and only AF relay with $P_R$ $=$ 33dBm. When $P_s$ $=$ 30dBm, the rate of LC-SCA-FP method is approximately 2.4 bits/s/Hz higher than those of three benchmark schemes.

Fig. 3 illustrates the achievable rate versus $P_i$ with $(M, N, K, P_s)$ $=$ (2, 32, 4, 30dBm). It is particularly noted that the proposed LC-SCA-FP method with $P_r$ $=$ 30dBm make a better rate performance improvement than that of AF relay+passive IRS, AF relay+passive IRS with random phase and only AF relay with $P_R=P_i+P_r$. For example, when $P_i$ equals 40dBm, the proposed LC-SCA-FP method has a 53.3\% rate gain over AF relay+passive IRS. This shows that as $P_i$ increases, significant rate gains are achieved for the proposed hybrid IRS-aided AF relay wireless network.

Fig. 4 presents the achievable rate versus the number $K$ of active IRS elements with $(M, N, P_s)$ $=$ (2, 32, 30dBm) for the proposed LC-SCA-FP method with $(P_i, P_r)$ $=$ (30dBm, 30dBm) and the three benchmark schemes with $P_R$ $=$ 33dBm. From Fig. 4, it can be observed that with the growth of the number $K$ of active IRS elements, the rate gain of LC-SCA-FP method over AF relay+passive IRS, AF relay+passive IRS with random phase and only AF relay increase gradually. Specifically, when $K=4$, the proposed LC-SCA-FP method can approximately harvest 60.0\% rate gain over the three benchmark schemes.

\section{Conclusions}
In this paper, we have proposed a low-complexity scheme called LC-SCA-FP to jointly optimize the beamforming matrix at AF relay and reflecting coefficient matrices at IRS in a hybrid IRS-aided AF relay network, where the hybrid IRS includes few active elements. Simulation results show that LC-SCA-FP can make a dramatic rate enhancement compared to AF relay+passive IRS, AF relay+passive IRS with random phase and only AF relay.

\ifCLASSOPTIONcaptionsoff
  \newpage
\fi

\bibliographystyle{IEEEtran}
\bibliography{IEEEfull,reference}
\end{document}